\documentclass[aps,pra,twocolumn,superscriptaddress,showpacs]{revtex4}
\def\be{\begin{equation}}
\def\ee{\end{equation}}
\def\bea{\begin{eqnarray}}
\def\eea{\end{eqnarray}}
\def\bi{\begin{itemize}}
\def\ei{\end{itemize}}

\usepackage{graphicx}
\usepackage{amsmath}
\usepackage{amssymb}

\begin{document}

\title{Entropy of entanglement and correlations induced by a quench:\\
          Dynamics of a quantum phase transition in the quantum Ising model}

\author{Lukasz Cincio}
\author{Jacek Dziarmaga}
\author{Marek M. Rams}
\affiliation{Institute of Physics and Centre for Complex Systems Research, 
             Jagiellonian University,
             Reymonta 4, 30-059 Krak\'ow, Poland}
\affiliation{Theory Division, Los Alamos National Laboratory, Los Alamos, NM 87545, USA}
\author{Wojciech H. Zurek}
\affiliation{Theory Division, Los Alamos National Laboratory, Los Alamos, NM 87545, USA}

\begin{abstract}
Quantum Ising model in one dimension is an exactly solvable example of a quantum phase transition. We investigate its behavior during a quench caused by a gradual turning off of the transverse bias field. The system is then driven at a fixed rate characterized by the quench time $\tau_Q$ across the critical 
point from a paramagnetic to ferromagnetic phase. In agreement with Kibble-Zurek mechanism 
(which recognizes that evolution is approximately adiabatic far away, but becomes approximately 
impulse sufficiently near the critical point), quantum state of the system after the transition exhibits 
a characteristic correlation length $\hat\xi$ proportional to the square root of the quench time 
$\tau_Q$: $\hat\xi=\sqrt{\tau_Q}$. The inverse of this correlation length is known to determine average density of defects (e.g. kinks) after the transition. In this paper, we show that this same $\hat\xi$ controls 
the entropy of entanglement, e.g. entropy of a block of $L$ spins that are entangled with the rest of 
the system after the transition from the paramagnetic ground state induced by the quench. For large 
$L$, this entropy saturates at $\frac16\log_2\hat\xi$, as might have been expected from the 
Kibble-Zurek mechanism. Close to the critical point, the entropy saturates when the block size 
$L\approx\hat\xi$, but -- in the subsequent evolution in the ferromagnetic phase -- a somewhat larger length scale $l=\sqrt{\tau_Q}\ln\tau_Q$ develops as a result of a dephasing process that can be regarded as a quantum analogue of phase ordering, and the entropy saturates when $L\approx l$. 
We also study the spin-spin correlation using both analytic methods and real time simulations with the Vidal algorithm. We find that at an 
instant when quench is crossing the critical point, ferromagnetic correlations decay exponentially with the dynamical correlation length $\hat\xi$, but (as for entropy of entanglement) in the following evolution length scale $l$ gradually develops. The correlation function becomes oscillatory at distances less than this scale. However, both the wavelength and the correlation length 
of these oscillations are still determined by $\hat\xi$. We also derive probability distribution for the number of kinks in a finite spin chain after the transition.

\pacs{ 03.65.-w, 73.43.Nq, 03.75.Lm, 32.80.Bx, 05.70.Fh }
\end{abstract}

\maketitle

\section{ Introduction }

Phase transition is a fundamental change in the character of the state of a system when
one of its parameters 
passes through 
the critical point. States on the opposite sides of the critical point are characterized by
different types of ordering. In a second order phase transition, the fundamental change 
is continuous and the critical point is characterized by divergences in the coherence
(or healing) length and in the relaxation time. This critical slowing down implies that no matter 
how slowly a system is driven through the transition its evolution cannot be adiabatic close 
to the critical point. If it were adiabatic, then the system would continuously evolve between 
the two types of ordering. However, in the wake of the necessarily non-adiabatic (and approximately 
impulse) evolution in the critical region, ordering of the state after the transition is not perfect: 
It is a mosaic of ordered domains whose finite size depends on the rate of the transition. 
This scenario was first described in the cosmological setting by Kibble \cite{K} who appealed
to relativistic casuality to set the size of the domains. The dynamical mechanism relevant for 
second order phase transitions was proposed by one of us \cite{Z}. It is based on the universality 
of critical slowing down, and leads to prediction that the size of the ordered domains scales with 
the transition time $\tau_Q$ as $\tau_Q^w$, where $w$ is a combination of critical exponents.
The~Kibble-Zurek mechanism (KZM) for second order thermodynamic phase transitions was 
confirmed by numerical simulations of the time-dependent Ginzburg-Landau model \cite{KZnum} 
and successfully tested by experiments in liquid crystals \cite{LC}, superfluid helium 3 \cite{He3},
both high-$T_c$  \cite{highTc} and low-$T_c$ \cite{lowTc} superconductors and even in 
non-equilibrium systems \cite{ne}. With the exception of superfluid $^4$He -- where the early detection 
of copious defect formation \cite{He4a} was subsequently attributed to vorticity inadvertently introduced by stirring \cite{He4b}, 
and the situation remains unclear -- experimental results are consistent with KZM, although more
experimental work is clearly needed to allow for more stringent experimental tests of KZM.  

The Kibble-Zurek mechanism is thus a universal theory of the dynamics
of second order phase transitions whose applications range from the low temperature
Bose-Einstein condensation (BEC) \cite{BEC} to the ultra high temperature transitions
in the grand unified theories of high energy physics. However, the zero temperature
quantum limit remained unexplored until very recently, see
Refs.\cite{3sites,KZIsing,Polkovnikov,Dziarmaga2005,Levitov}
and an example of disordered system in Ref. \cite{random}, and
quantum phase transitions are in many respects qualitatively different from transitions
at finite temperature. Most importantly time evolution is unitary, so there is no damping,
and there are no thermal fluctuations that initiate symmetry breaking in KZM.

Quantum state of the many body system is indeed profoundly different from a classical state: Instead 
of a single broken symmetry configuration it may (and, generally, will) contain {\it all} of the possible 
configurations in a superposition. In addition to the `classical' way of characterizing this state through 
the density of excitations (e.g. defects), one can wonder how entangled various parts of the system are
with each other. Von Neumann entropy of a fragment due to its entanglement with the rest of the system is 
a convenient way to quantify this. It can be computed as a function of the size of the fragment. In equilibrium, 
and away from the critical point, this entropy of entanglement saturates at distances of the order of the 
coherence length $\xi$ of the system at values $\sim \ln\xi$ for one dimensional systems \cite{18a}. However, 
at the critical point (where equilibrium coherence length $\xi$ becomes
infinite) entropy of entanglement diverges with the size of the fragment. In particular, in one dimensional 
systems, the critical entanglement 
entropy diverges logarithmically ($\sim \ln L$) with the length $L$ of the chain fragment \cite{17a,17b,18a}.

This equilibrium behavior suggests the question: What is the entanglement entropy left behind by 
an out-of-equilibrium phase transition? Such a transition will pass through the critical point 
(where entanglement entropy is logarithmically divergent) but this will happen at a finite
rate set by the quench time $\tau_Q$. We show that the resulting entanglement entropy is of the order 
of $\sim \ln \hat \xi$, where $\hat \xi$ is the healing (coherence) length at the instant  when critical 
slowing down forces the system to switch from the approximately adiabatic to approximately impulse (`diabatic') 
behavior. This suggests that the same process that determines the size of regions that ``break symmetry in 
unison'' (which sets the density of topological defects left by the transition) is also 
responsible for the resulting entanglement of formation left by the quench. This finding is consistent with 
recent results on quantum phase transitions induced by instantaneous quenches \cite{18b,18c,Instant} which 
indicate that structures present in the initial pre-transition state determine the structures (and hence 
entanglement of 
formation) that arise after an instantaneous quench: Our results also suggest that 
-- in accord with KZM -- it is a good approximation to consider quench to be approximately adiabatic until 
the instant $\hat t \sim 1/\sqrt{\tau_Q}$ before the critical point is reached, and approximately impulse 
(e.g. nearly instantaneous) inside this time interval. This also confirms and extends results of the recent
study of Cherng and Levitov \cite{Levitov} who computed entropy density and correlations induced
by quenches in one-dimensional chains, and concluded that their results support KZM.

While our results below are established for the one-dimensional quantum Ising model (which has the
considerable advantage of being exactly solvable), we conjecture that similar behavior will be encountered in other quantum phase transitions, and that their non-equilibrium evolution can be
anticipated using equilibrium critical exponents using KZM. This conjecture can be then tested in 
a variety of systems that undergo quantum phase transitions both in condensed matter and in
atomic physics experiments.

According to Sachdev \cite{book}, the understanding of quantum phase transitions is based
on two prototypical models. One is
the quantum rotor model 
and the other is the one-dimensional quantum Ising model. Of the
two only the Ising model is exactly solvable. It is defined by the Hamiltonian
\be
H~=~-\sum_{n=1}^N \left( g\sigma^x_n + \sigma^z_n\sigma^z_{n+1} \right)~.
\label{Hsigma}
\ee
with periodic boundary conditions
\be
\vec\sigma_{N+1}~=~\vec\sigma_1~.
\label{periodicsigma}
\ee
Quantum phase transition takes place at the
critical value $g=1$ of an external magnetic field. When $g\gg 1$, the ground
state is a paramagnet
$|\rightarrow\rightarrow\rightarrow\dots\rightarrow\rangle$ with all
spins polarized along the $x$-axis. On the other hand, when $g\ll 1$, then
there are two degenerate ferromagnetic ground states with all spins
pointing either up or down along the $z$-axis:
$|\uparrow\uparrow\uparrow\dots\uparrow\rangle$ or
$|\downarrow\downarrow\downarrow\dots\downarrow\rangle$. In an infinitesimally slow classical
transition from paramagnet to ferromagnet, the system would choose one of the
two ferromagnetic states. In the analogous quantum case, any superposition of these two states
is also a `legal'  ground state providing it is consistent with other quantum numbers (e.g. parity)
conserved by the transition from the initial paramagnetic state.  
However, when $N\to\infty$, then energy gap at $g=1$ tends to zero (quantum version of the
critical slowing down) and it is
impossible to pass the critical point at a finite speed without exciting the system.
As a result, the system ends in a quantum superposition of states like
\be
|\dots
\uparrow
\downarrow\downarrow\downarrow\downarrow\downarrow
\uparrow\uparrow\uparrow\uparrow\uparrow\uparrow\uparrow
\downarrow\downarrow\downarrow\downarrow
\uparrow\uparrow\uparrow\uparrow\uparrow\uparrow
\downarrow
\dots\rangle
\label{domains}
\ee
with finite domains of spins pointing up or down and separated by kinks where the
polarization of spins changes its orientation. Average size of the domains or,
equivalently, average density of kinks depends on a transition rate. When
the transition is slow, then the domain size is large, but when it is very fast, then orientation
of individual spins can become random, uncorrelated with their nearest neighbors.
Transition time $\tau_Q$ can be unambiguously defined when we assume that close to
the critical point at $g=1$ time-dependent field $g(t)$ driving the transition can
be approximated by a linear quench
\be
g(t<0)~=~-\frac{t}{\tau_Q}~.
\label{linear}
\ee
with the adjustable quench time $\tau_Q$. Density of kinks after the linear quench was
estimated in Ref.~\cite{KZIsing} showing that KZM can be also applied to quantum phase 
transitions. In this derivation, it is convenient to use instead of $g(t)$ a dimensionless parameter 
$\epsilon(t)=\frac{g-g_c}{g_c}=g-1$. As in classical transitions \cite{Z}, one can assume
the adiabatic-impulse approximation \cite{Bodzio,Bodzio2}. The quench begins in the ground 
state at large initial $g$ and the initial part of the evolution is adiabatic:
the state follows the instantaneous ground state of the system. The evolution
becomes non-adiabatic close to the critical point when the energy gap 
$\simeq|\epsilon|$ becomes comparable to the instantaneous transition rate
$|\dot \epsilon/\epsilon|$. This condition leads to an equation solved by $\hat t = \sqrt \tau_Q$ --
the instant when the adiabatic to impulse transition occurs --  which in turn yields
 $\hat\epsilon\simeq\tau_Q^{-1/2}$ and corresponds to the 
coherence length in the ground state \cite{Z, KZIsing}:
\be
\hat\xi~=~\tau_Q^{1/2}~.
\ee
Assuming impulse approximation, the quantum state does not change during
the following non-adiabatic stage of the evolution between $\hat\epsilon$ 
and $-\hat\epsilon$. Consequently, the quantum state at $-\hat\epsilon$
is expected to be approximately the ground state at $\hat\epsilon$ with the coherence length
proportional to $\hat\xi$ and this is the initial state for the final
adiabatic stage of the evolution after $\hat\epsilon$. This argument
shows that when passing across the critical point, the state of the system
gets imprinted with a finite KZ correlation length proportional to $\hat\xi$. 
In particular, this coherence length determines average density of kinks
after the transition as 
\be
n~\simeq~\frac{1}{\tau_Q^{1/2}}~.
\label{KZscaling}
\ee
This is an order of magnitude estimate with an unknown ${\cal O}(1)$ prefactor. 
The estimate was verified by numerical simulations in Ref.~\cite{KZIsing}.
Not much later the problem was solved exactly in Ref.~\cite{Dziarmaga2005}, see also
Ref.~\cite{Levitov}, with the exact solution confirming the KZM scaling in
Eq.~(\ref{KZscaling}). 

In the next section we review and expand the exact solution, and then
use the expanded version to obtain a more complete set of results. In Subsection
\ref{probability}, we derive Gaussian probability distribution for the number 
of kinks measured after a quench in a finite Ising spin chain. In Section 
\ref{entropy}, we calculate entropy of entanglement of a block of $L$ spins
after a dynamical transition and in Section \ref{correlations} we work out 
spin-spin correlation functions. We conclude in Section \ref{conclusion}.

\section{ Exact solution }\label{exact}
\subsection{ Energy spectrum }

Here we assume that the number of spins $N$ is even for convenience. After the nonlocal
Jordan-Wigner transformation \cite{JW},
\bea
&&
\sigma^x_n~=~1-2 c^\dagger_n  c_n~, \\
&&
\sigma^z_n~=~
-\left( c_n+ c_n^\dagger\right)
 \prod_{m<n}(1-2 c^\dagger_m c_m)~,
\eea
introducing fermionic operators $c_n$ which satisfy anticommutation relations 
$\left\{c_m,c_n^\dagger\right\}=\delta_{mn}$ and 
$\left\{ c_m, c_n \right\}=\left\{c_m^\dagger,c_n^\dagger \right\}=0$
the Hamiltonian (\ref{Hsigma}) becomes \cite{LSM}
\be
 H~=~P^+~H^+~P^+~+~P^-~H^-~P^-~.
\label{Hc}
\ee
Above;
\be
P^{\pm}=
\frac12\left[1\pm\prod_{n=1}^N\sigma^x_n\right]=
\frac12\left[1~\pm~\prod_{n=1}^N\left(1-2c_n^\dagger c_n\right)\right]
\label{Ppm}
\ee
are projectors on the subspaces with even ($+$) and odd ($-$) numbers of 
$c$-quasiparticles and  
\bea
H^{\pm}~=~
\sum_{n=1}^N
\left( 
g c_n^\dagger  c_n - c_n^\dagger  c_{n+1} - c_{n+1}  c_n - \frac{g}{2} +{\rm h.c.} 
\right)~.
\label{Hpm}
\eea
are the corresponding reduced Hamiltonians. The $c_n$'s in $H^-$ satisfy
periodic boundary conditions $c_{N+1}=c_1$, but the $c_n$'s in $H^+$
must obey $c_{N+1}=-c_1$ -- e.g,  ``antiperiodic'' boundary conditions. 

The parity of the number of $c$-quasiparticles is a good quantum number and the ground
state has even parity for any non-zero value of $g$. Assuming that a quench begins in 
the ground state, we can confine to the subspace of even parity. $H^+$ is diagonalized 
by the Fourier transform followed by Bogoliubov transformation \cite{LSM}. The Fourier transform
consistent with the antiperiodic boundary condition $c_{N+1}=-c_1$ is  
\be
c_n~=~ 
\frac{e^{-i\pi/4}}{\sqrt{N}}
\sum_k c_k e^{ikn}~,
\label{Fourier}
\ee
where pseudomomenta $k$ take ``half-integer'' values:
\be
k~=~
\pm \frac12 \frac{2\pi}{N},
\dots,
\pm \frac{N-1}{2} \frac{2\pi}{N}~.
\label{k}
\ee
It transforms the Hamiltonian into
\bea
H^+~&=&
\sum_k
\left\{
2[g-\cos(ka)] c_k^\dagger c_k + \right. \nonumber\\
&&
\left. \sin(ka)
\left[ 
 c^\dagger_k c^\dagger_{-k}+
 c_{-k} c_k
\right]
-g
\right\}~.
\label{Hck}
\eea
Diagonalization of $H^+$ is completed by the Bogoliubov transformation
\be
c_k~=~
u_k  \gamma_k + v_{-k}^*  \gamma^\dagger_{-k}~,
\label{Bog}
\ee
provided that Bogoliubov modes $(u_k,v_k)$ are eigenstates of the stationary
Bogoliubov-de Gennes equations
\bea
\epsilon~ u_k &=& +2[g-\cos k] u_k+2\sin k v_k~,\nonumber\\
\epsilon~ v_k &=& -2[g-\cos k] v_k+2\sin k u_k~. \label{stBdG}
\eea
There are two eigenstates for each $k$ with eigenenergies $\epsilon=\pm\epsilon_k$,
where  
\be
\epsilon_k~=~2\sqrt{[g-\cos k]^2+\sin^2 k}~.
\label{epsilonk}
\ee
The positive energy eigenstate 
\be
(u_k,v_k)\sim\left[(g-\cos k)+\sqrt{g^2-2g\cos k+1},\sin k\right]~,
\label{uvplus}
\ee
which has to be normalized so that $|u_k|^2+|v_k|^2=1$, defines the quasiparticle 
operator $\gamma_k~=~u_k^* c_k + v_{-k} c_{-k}^\dagger$,
and the negative energy eigenstate 
$
(u^-_k,v^-_k)=(-v_k,u_k)~ 
\label{uv-}
$
defines $\gamma_k^-=(u_k^-)^*c_k+v_{-k}^-c_{-k}^\dagger=-\gamma_{-k}^\dagger$. 
After the Bogoliubov transformation, the Hamiltonian 
$
H^+=
\frac12
\sum_k \epsilon_k
\left( 
\gamma_k^\dagger \gamma_k - \gamma_k^{-\dagger} \gamma_k^-    
\right)
$
is equivalent to
\be
H^+~=~
\sum_k \epsilon_k~
\left(
\gamma_k^\dagger \gamma_k-\frac12
\right)~.
\label{Hgamma}
\ee
This is a simple-looking sum of quasiparticles with half-integer pseudomomenta. 
However, thanks to the projection $P^+~H^+~P^+$ in Eq.~(\ref{Hc}) only states 
with even numbers of quasiparticles belong to the spectrum of $H$.   

\subsection{ Linear quench }

In the linear quench Eq.~(\ref{linear}), the system is initially ($t\ll -\tau_Q$) in its 
ground state at large initial value of $g\gg 1$, but when $g$ is ramped down to 
zero, the system gets excited from its instantaneous ground state
and, in general, its final state at $t=0$ has finite number of kinks. Comparing 
the Ising Hamiltonian Eq.~(\ref{Hsigma}) at $g=0$ with the Bogoliubov Hamiltonian 
(\ref{Hgamma}) at $g=0$ we obtain a simple expression for the operator of the 
number of kinks 
\bea
{\cal N} ~\equiv~ \frac12 \sum_{n=1}^{N} 
                  \left(1-\sigma^z_n\sigma^z_{n+1}\right)~=~
                  \sum_k \gamma_k^\dagger \gamma_k~.   
\label{calN}
\eea 
The number of kinks is equal to the number of quasiparticles excited at $g=0$. 
The excitation probability 
\be
p_k~=~\langle\psi(0)|\gamma_k^\dagger\gamma_k|\psi(0)\rangle~
\ee
in the final state can be found with the time-dependent Bogoliubov method. 

The initial ground state is Bogoliubov vacuum $|0\rangle$ annihilated by all 
quasiparticle operators $\gamma_k$ which are determined by the asymptotic form 
of the (positive energy) Bogoliubov modes $(u_k,v_k)\approx(1,0)$ in the regime 
of $g\gg1$. When $g(t)$ is ramped down, the quantum state $|\psi(t)\rangle$
gets excited from the instantaneous ground state. The time-dependent 
Bogoliubov method makes an Ansatz that $|\psi(t)\rangle$ is a Bogoliubov vacuum 
annihilated by a set of quasiparticle annihilation operators $\tilde{\gamma}_k$ 
defined by a time-dependent Bogoliubov transformation
\be
c_k ~=~ u_k(t)  \tilde{\gamma}_k + v_{-k}^*(t)  \tilde{\gamma}^\dagger_{-k}~,
\label{tildeBog}
\ee
with the initial condition $[u_k(-\infty),v_k(-\infty)]=(1,0)$. In the Heisenberg
picture, the Bogoliubov modes $[u_k(t),v_k(t)]$ must satisfy Heisenberg equation
$
i\hbar\frac{d}{dt} c_k~=~\left[ c_k, H^+\right]
$
with the constraint that $\frac{d}{dt}\tilde{\gamma}_k=0$. The Heisenberg equation 
is equivalent to the dynamical version of the Bogoliubov-de Gennes 
equations (\ref{stBdG}):
\bea
i\hbar\frac{d}{dt} u_k &=&
+2\left[g(t)-\cos k\right] u_k +
 2 \sin k~ v_k~,\nonumber\\
i\hbar\frac{d}{dt} v_k &=&
-2\left[g(t)-\cos k\right] v_k +
 2 \sin k~ u_k~.
\label{dynBdG} 
\eea
At any value of $g$, Eqs.~(\ref{dynBdG}) have two instantaneous eigenstates. 
Initially, the mode $[u_k(t),v_k(t)]$ is the positive energy eigenstate, but during 
the quench it gets ``excited'' to a combination of the positive and negative mode. 
At the end of the quench at $t=0$ when $g=0$ we have
\be
\left[u_k(0),v_k(0)\right]=
A_k
\left(u_k,v_k\right)+
B_k
\left(u_k^-,v_k^-\right)
\ee
and consequently $\tilde{\gamma}_k=A_k\gamma_k-B_k\gamma_k^\dagger$.
The final state which is, by construction, annihilated by both $\tilde{\gamma}_k$ and 
$\tilde{\gamma}_{-k}$ is
\be
|\psi(0)\rangle=
\prod_{k>0}
\left(
A_k+B_k\gamma^\dagger_k\gamma^\dagger_{-k}
\right)
|0\rangle~.
\label{finalstate}
\ee
Pairs of quasiparticles with pseudomomenta $(k,-k)$ are excited with probability
\be
p_k~=~|B_k|^2~,
\label{pbeta}
\ee
which can be found by mapping Eqs.~(\ref{dynBdG}) to the Landau-Zener (LZ) problem 
(similarity between KZM and LZ problem was first pointed out by Damski in 
Ref.~\cite{Bodzio}). 

The transformation 
\be
\tau~=~4\tau_Q\sin k\left(-\frac{t}{\tau_Q}+\cos k\right)
\label{tau}
\ee
brings Eqs.~(\ref{dynBdG}) to the standard LZ form \cite{LZF}
\bea
i\hbar\frac{d}{d\tau} u_k &=&
-\frac12(\tau\Delta_k) u_k + \frac12 v_k ~,\nonumber\\
i\hbar\frac{d}{d\tau} v_k &=&
+\frac12(\tau\Delta_k) v_k + \frac12 u_k ~,\label{LZ}
\label{BdGLZ}
\eea
with $\Delta_k^{-1}=4\tau_Q\sin^2 k$. Here the time $\tau$ runs from $-\infty$ to  
$\tau_{\rm final}=2\tau_Q\sin(2k)$ corresponding to $t=0$. Tunneling between the 
positive and negative energy eigenstates happens when 
$\tau\in(-\Delta_k^{-1},\Delta_k^{-1})$. $\tau_{\rm final}$ is well outside this 
interval, $\tau_{\rm final}\gg\Delta_k^{-1}$, for 
long wavelength modes with $|k|\ll\frac{\pi}{4}$. For these modes, time $\tau$ 
in Eqs.~(\ref{LZ}) can be extended to $+\infty$ making them fully equivalent to 
LZ equations \cite{LZF}. This equivalence can be used to easily obtain several
simple results \cite{Dziarmaga2005} described in the next subsection.

\subsection{ Simple results }

In the limit of slow transitions we can assume that only long wavelength modes, which
have small gaps at their anti-crossing points, can get excited. For these modes, 
we can use the LZ formula \cite{LZF} for excitation probability:
\be
p_k~\simeq~
e^{-\frac{\pi}{2\Delta_k}}~\approx~
e^{-2\pi\tau_Qk^2}~.
\label{LZpk}
\ee 
This approximation is self-consistent only when the width of the obtained
Gaussian $(4\pi\tau_Q)^{-1/2}$ is much less than $\frac{\pi}{4}$
or, equivalently, for slow enough quenches with $\tau_Q\gg 1$.
With the LZ formula (\ref{LZpk}), we can calculate the number of kinks 
in Eq.~(\ref{calN}) as
\be
{\cal N}~=~\sum_k~p_k~.
\label{LZn}
\ee
There are at least two interesting special cases:

\bi

\item When $N\to\infty$ the sum in 
Eq.~(\ref{LZn}) can be replaced by an integral. The expectation value of density of 
kinks becomes
\be
n=\lim_{N\to\infty}\frac{\cal N}{N}=
\frac{1}{2\pi}\int_{-\pi}^{\pi}dk~p_k=
\frac{1}{2\pi}
\frac{1}{\sqrt{2\tau_Q}}.
\label{scaling}
\ee    
The density scales like $\tau_Q^{-1/2}$ in agreement with KZM, see 
Eq.~(\ref{KZscaling}). Thus, slower quenches lead to fewer defects.

\item Following Ref.~\cite{KZIsing}, we can ask what the fastest $\tau_Q$ is when 
no kinks get excited in a finite chain of size $N$. This critical $\tau_Q$ marks 
a crossover between adiabatic and non-adiabatic regimes. In other words, we can 
ask what is the probability for a finite chain is to stay in the ground state. As 
different pairs of quasiparticles $(k,-k)$ evolve independently, the probability 
to stay in the ground state is the product
\be
{\cal P}_{\rm GS}~=~
\prod_{k>0}\left(1-p_k\right)~.
\label{calP}
\ee
Well on the adiabatic side only the pair $\left(\frac{\pi}{N},-\frac{\pi}{N}\right)$
is likely to get excited and we can approximate
\be
{\cal P}_{\rm GS}~\approx~
1-p_{\frac{\pi}{N}}~\approx~
1-\exp\left(-2\pi^3\frac{\tau_Q}{N^2}\right)~.
\label{calPapprox}
\ee
A quench in a finite chain is adiabatic when
\be
\tau_Q~\gg~\frac{N^2}{2\pi^3}~.
\label{nonadiabatic}
\ee
Reading this inequality from right to left, the size $N$ of a defect-free chain grows 
like $\tau_Q^{1/2}$. This is consistent with Eq.~(\ref{KZscaling},\ref{scaling}).

\ei

\subsection{ Probability distribution of the number of kinks }\label{probability}

Equation (\ref{scaling}) gives an average density of kinks measured after a quench
to zero magnetic field $g=0$. In a finite chain, an average number of kinks is 
$\overline{\cal N}=Nn$, provided that the transition is non-adiabatic unlike 
in Eq.~(\ref{nonadiabatic}). The average $\overline{\cal N}$ is an expectation
value of a probability distribution $P({\cal N})$ for the number of kinks
${\cal N}$ measured after a quench. 

The number of kinks ${\cal N}$ is the number of quasiparticles excited by the 
end of the quench. As quasiparticles are excited in pairs with opposite quasimomenta 
$(k,-k)$, the number of kinks ${\cal N}$ must be even. A pair $(k,-k)$ is excited with
the probability $p_k$ in Eq.~(\ref{LZpk}). We can asign to each pair of quasiparticles
a random variable $x_k$ which is $1$ when the pair is excited and $0$ otherwise: 
$x_k=1$ with probability $p_k$ and $x_k=0$ with probability $1-p_k$. We want a probability 
distribution for the sum ${\cal N}=\sum_{k>0}2x_k$. This is a sum of independent random 
variables of finite variance so for a large number of variables, the sum ${\cal N}$ becomes 
a Gaussian random variable with a mean $\overline{\cal N}=Nn$ and finite variance 
$\sim N$.

We have to be careful to specify when the number of random variables is large. Naively, 
on a $N$-site lattice, there are $N/2$ pairs of quasiparticles $(k,-k)$, or independent random 
variables $x_k$, and the number seems to be large when $N$ is large. On second thought, it 
is clear that variables with $p_k\approx 0$ or $p_k\approx 1$ cannot really count because 
they are hardly random at all. A look at the Gaussian $p_k$ in Eq.~(\ref{LZpk}) shows that 
the range of $k>0$ where 
$0 \ll p_k \ll 1$ has width $\simeq \frac{1}{\sqrt{\tau_Q}}$ which accommodates 
$\simeq \frac{N}{\sqrt{\tau_Q}}$ discrete values of pseudomomentum $k$. The relevant
number of random variables is $\simeq \frac{N}{\sqrt{\tau_Q}}\simeq N n=\overline{\cal N}$.
It is large when the average number of kinks is large,
\be
\overline{\cal N}~\gg~1~.
\ee 
With this assumption $P(\cal N)$ is Gaussian. 

Keeping this assumption in mind we can proceed as
\bea
P({\cal N})&=&
\sum_{x_k}~
\delta_{{\cal N},\sum_{k>0}2x_k} 
\nonumber\\
&&
\prod_{k>0}
\left[
\delta_{0,x_k}(1-p_k)+\delta_{1,x_k}p_k
\right]~= 
\label{Pprod}\\
&&
\frac{1}{2\pi}\int_{-\pi}^\pi dq e^{-iq{\cal N}}
\prod_{k>0}
\left[
(1-p_k)+p_k e^{2iq}
\right]~\equiv
\nonumber\\
&&
\frac{1}{2\pi}\int_{-\pi}^\pi dq e^{-iq{\cal N}}
\tilde{P}(q)~.
\label{tildeP}
\eea
It is convenient to evaluate first
\bea
\ln\tilde{P}(q)&=&
\sum_{k>0} \ln\left[(1-p_k)+p_k e^{2iq}\right]~\approx
\nonumber\\
&&
\frac{N}{2\pi}\int_0^\pi dk~\ln\left[(1-p_k)+p_k e^{2iq}\right]~.
\eea
Here we used $N\gg 1$, a necessary condition for our assumption that 
$\bar{\cal N}\gg 1$. After changing the integration variable to 
$u=\frac{k}{2\pi n}$, we can extend the integration over $u$ to infinity:
\bea
\ln\tilde{P}(q)&=&
\overline{{\cal N}}
\int_0^\infty du~
\ln\left[(1-e^{-\pi u^2})+e^{-\pi u^2+2iq}\right]~\approx
\nonumber\\
&&
\overline{{\cal N}}~
\left(
i q - \frac{\sqrt{2}-1}{\sqrt{2}} q^2
\right)+
{\cal O}(q^3)
~.
\label{q3}
\eea 
Here we used again $\overline{{\cal N}}=Nn$. 

Finally, combination of Eqs. (\ref{q3},~\ref{tildeP}) gives the expected
Gaussian probability distribution
\bea
P\left( {\cal N} \right)&=&
\frac{1}{\sqrt{2\pi\sigma^2_{\cal N}}}
\exp
\left(
-\frac{({\cal N}-\overline{\cal N})^2}{2\sigma^2_{\cal N}}
\right)~.
\label{PcalNgauss}
\eea
with a variance
\be
\sigma^2_{\cal N}~=~
\left(2-\sqrt{2}\right)~\overline{\cal N}~.
\label{variancecalN}
\ee
In the course of our approximations, we have lost the constraint that 
${\cal N}$ must be even, but this is not a big mistake when
$\overline{{\cal N}}\gg 1$.   

It is also interesting to study the opposite adiabatic regime when $Nn\ll 1$. 
When $\tau_Q$ is large we can approximate the product in Eq.~(\ref{Pprod})
by a single factor with the lowest $k=\pi/N$, 
\bea
P({\cal N})&\approx&
\sum_{s=0,1}
\delta_{{\cal N},2s}~ 
\left[
\delta_{0,s}\left(1-p_{\pi/N}\right)+\delta_{1,s}p_{\pi/N}
\right]~=
\nonumber\\
&&
\delta_{{\cal N},0}~\left(1-p_{\pi/N}\right)~+~
\delta_{{\cal N},2}~p_{\pi/N}~.
\eea
This is a good approximation when $\tau_Q~\gg~\frac{N^2}{2\pi^3}$ as in 
Eq.~(\ref{nonadiabatic}). In this adiabatic regime 
\be
\overline{\cal N}~=~
2 p_{\pi/N}~=~
\exp\left(-2\pi^3\frac{\tau_Q}{N^2}\right)~.
\ee
The average number of kinks decays exponentially with $\tau_Q$ and the 
KZ power law scaling (\ref{scaling}) does not extend to this 
adiabatic regime, as was already noted in Ref.~\cite{KZIsing}.

\subsection{ Exact solution and the two scales of length }\label{twoscales}

So far we have avoided writing down solutions of the Landau-Zener equations (\ref{BdGLZ})
whose general form is, see e.g. Appendix B in Ref.~\cite{Bodzio2}, 
\bea
v_k(\tau)&=&
-\left[ a D_{-s-1}(-iz) + b D_{-s-1}(iz) \right] ~, \nonumber\\
u_k(\tau)&=&
\left(-\Delta_k\tau+2i\frac{\partial}{\partial\tau}\right)
v_k(\tau) ~,
\label{general} 
\eea
with arbitrary complex parameters $a,b$. Here $D_m(x)$ is a Weber function, 
$s=\frac{1}{4i\Delta_k}$, and $z=\sqrt{\Delta_k}\tau e^{i\pi/4}$.
The parameters $a,b$ are fixed by the initial conditions 
$u_k(-\infty)=1$ and $v_k(-\infty)=0$. Using the asymptotes of the Weber
functions when $\tau\to-\infty$, we get $a=0$ and 
\be
|b|^2=\frac{e^{-\pi/8\Delta_k}}{4\Delta_k}~.
\ee 
The solution of the linear quench problem is then
\bea
v_k(\tau)&=&
- b D_{-s-1}(iz) ~, \nonumber\\
u_k(\tau)&=&
\left(-\Delta_k\tau+2i\frac{\partial}{\partial\tau}\right)
v_k(\tau) ~,
\label{solution} 
\eea
At the end of the quench for
$t=0$ and when $\tau=\tau_k=2\tau_Q \sin(2k)$, the argument of the Weber function
$iz=\sqrt{\Delta_k}\tau e^{i\pi/4}=2\sqrt{\tau_Q}e^{i\pi/4}\cos(k){\rm sign}(k)$.
In the limit of large $\tau_Q$ the modulus of this argument is large 
for most $k$, except the neighborhoods of $k=\pm\frac{\pi}{2}$, and we can 
again use the asymptotes of the Weber functions. After some work
we get the products
\bea
|u_k|^2&=&
\frac{1-\cos k}{2}+
e^{-2\pi\tau_Q\sin^2 k}~,\nonumber\\
|v_k|^2&=&
1-|u_k|^2~,\nonumber\\
u_k v_k^* &=&
\frac12\sin k +\nonumber\\
&&
{\rm sign}(k)~
e^{-\pi\tau_Q\sin^2 k}
\sqrt{1-e^{-\pi\tau_Q\sin^2 k}}~
e^{i\varphi_k}~,\nonumber\\
\varphi_k &=&
\frac{\pi}{4}+
\frac{\Delta_k \tau_k^2}{2}+
\frac{\ln\Delta_k}{4\Delta_k}+
\frac{\ln\tau_k}{2\Delta_k}-   \nonumber\\
&&
\arg\left[\Gamma\left(1+\frac{i}{4\Delta_k}\right)\right]~.
\label{att0}
\eea
Here $\Gamma(x)$ is the gamma function.

We expect that for large $\tau_Q$ only modes with small $|k|\ll\frac{\pi}{4}$ 
get excited. In this long wave length limit, the products can be further
simplified to
\bea
|u_k|^2&=&
\frac{1-\cos k}{2}+
e^{-2\pi\tau_Q k^2}~,\nonumber\\
|v_k|^2&=&
1-|u_k|^2~,\nonumber\\
u_k v_k^* &=&
\frac12\sin k+
{\rm sign}(k)~
e^{-\pi\tau_Q k^2}
\sqrt{1-e^{-\pi\tau_Q k^2}}~
e^{i\varphi_k}~,\nonumber\\
\varphi_k &=&
\frac{\pi}{4}+
2\tau_Q-
(2-\ln 4)\tau_Q k^2+ \nonumber\\
&&
k^2 \tau_Q\ln\tau_Q-
\arg\left[\Gamma\left(1+i\tau_Q k^2\right)\right]~.
\label{att0smallk}
\eea
These products depend on $k$ and $\tau_Q$ through two combinations:
$\tau_Q k^2$, which implies the usual KZM coherence length
$\hat\xi=\sqrt{\tau_Q}$, and $k^2 \tau_Q\ln\tau_Q$ which implies
a second length scale $\sqrt{\tau_Q\ln\tau_Q}$. The final quantum state
at $g=0$ cannot be characterized by a single scale of length. Physically, this
appears to reflect a combination of two processes: KZM that sets up initial 
post-transition state of the system, and the subsequent evolution that can be
regarded as quantum phase ordering.

\section{Entropy of a block of spins}\label{entropy}

Von Neumann entropy of a block of $L$ spins due to its entanglement with the rest of the system;
\be
S(L)~=~-~{\rm Tr}~\rho_L~\log_2\rho_L~,
\ee
is a convenient measure of entanglement.
Above $\rho_L$ is reduced density matrix of the subsystem of $L$ spins. In recent
years this entropy was studied extensively in ground states of quantum critical 
systems \cite{17b,17c,17d,17e,18a,18b,18c,JStatPhys}. At a quantum critical point, the entropy 
diverges like $\log L$ for large $L$ with a prefactor determined by the central charge 
of the relevant conformal field theory \cite{17a,17b,18a}. In particular, in the quantum Ising 
model at the critical $g=1$
\be
S^{\rm GS}(L) ~\simeq~ \frac16 \log_2 L~
\label{SGCcrit}
\ee 
for large $L$. Slightly away from the critical point, the entropy saturates at
a finite asymptotic value \cite{18a}
\be
S_{\infty}^{\rm GS} ~\simeq~ \frac16 \log_2 \xi 
\label{SGSoff}
\ee 
when the block size $L$ exceeds the finite correlation length $\xi$ in the
ground state of the system.

\subsection{ Entropy after dynamical transition }

In a dynamical quantum phase transition the quantum state of the system
developes a finite correlation length $\hat\xi\simeq\sqrt{\tau_Q}$. If this
dynamical correlation length were the only relevant scale of length, then 
one could expect entropy of entanglement after a dynamical transition given by 
Eq.~(\ref{SGSoff}) with $\xi$ simply replaced by $\hat\xi$. However, as we 
saw in Eq.~(\ref{att0smallk}), there are two scales of length, and -- strictly speaking -- there 
is no reason to expect that either of them alone is relevant in general. This
is why we shall not rely on scaling arguments alone and will go on to calculate the entropy 
of entanglement ``from scratch''.

We proceed in a similar way as in Ref. \cite{17b,17c,17e,18b,JStatPhys} and
define a correlator matrix for the block of $L$ spins 
\be
\Pi~=~
\left(
\begin{array}{ccc}
\alpha &,& \beta^{\dagger} \\  
\beta  &,& 1-\alpha  
\end{array}
\right)~,
\label{rhoblock}
\ee
where $\alpha$ and $\beta$ are $L\times L$ matrices of quadratic correlators
\bea
\alpha_{m,n} &\equiv& \langle c_m c_n^\dagger \rangle = \nonumber\\
&& \frac{1}{2\pi}\int_{-\pi}^\pi dk~|u_k|^2~e^{ik(m-n)}~
   \stackrel{\tau_Q\gg 1}{\approx}   \nonumber\\
&& \frac12\delta_{0,|m-n|}-\frac14\delta_{1,|m-n|}+
   \frac{e^{-\frac{(m-n)^2}{8\pi~\hat\xi^2}}}{2\sqrt{2}\pi~\hat\xi}~.
\label{alpha}
\eea
and 
\bea
&& \beta_{m,n}~\equiv~\langle c_m c_n \rangle = \nonumber\\
&& \frac{1}{2\pi i}\int_{-\pi}^\pi dk~u_kv_k^*~e^{ik(m-n)}~= \nonumber\\
&& \frac{1}{\pi}\int_0^\pi dk~u_kv_k^*~\sin k(m-n) ~
   \stackrel{\ln\tau_Q\gg 1}{\approx}
   \label{beta}\\
&& {\rm sign}(m-n)~\times \nonumber\\
&& \left[
   \frac14 \delta_{1,|m-n|}-
   \frac{e^{i\left(2\tau_Q-\frac{|m-n|^2}{4~\hat\xi l}\right)}}
        {2\sqrt{\pi~\hat\xi~l}}
   e^{-\frac{\pi|m-n|^2}{4~l^2}}
   \sqrt{1-e^{-\frac{\pi|m-n|^2}{4~l^2}}}
   \right] 
\nonumber
\eea
Here $\hat\xi=\sqrt{\tau_Q}$ is the KZM dynamical correlation length and
\be 
l~\equiv~\sqrt{\tau_Q}\ln\tau_Q~. 
\label{l}
\ee
We note that $\alpha$ and $\beta$ are Toeplitz matrices with constant diagonals. 
The expectation values $\langle\dots\rangle$ are taken in the dynamical
Bogoliubov vacuum state. As this state is Gaussian, all higher order correlators
can be expressed by the matrices $\alpha$ and $\beta$ - they provide complete
characterization of the quantum state after the dynamical transition.
The matrices depend on both scales $\hat\xi$ and $l$ and both scales are
necessary to characterize the Gaussian state.

\begin{figure}[t]
\includegraphics[width=\columnwidth,clip=true]{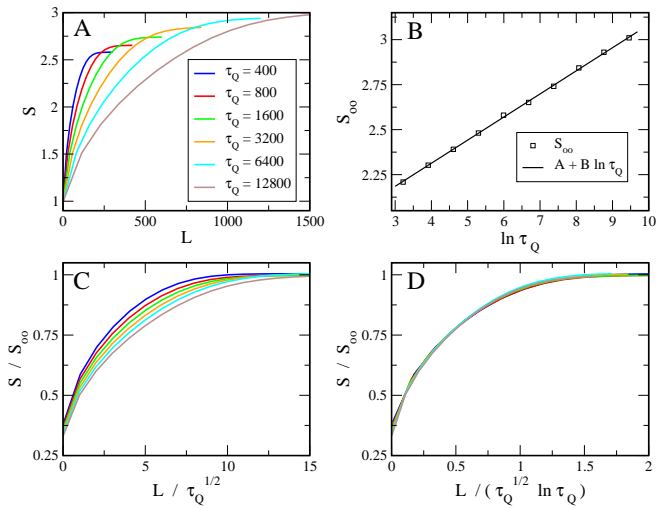}
\caption{
Panel A shows entropy of a block of $L$ spins after the dynamical phase
transition as a function of the block size $L$. The multiple plots
correspond to different values of the quench time $\tau_Q$. For all $\tau_Q$,
the entropy grows with the block size $L$ and saturates at a
finite value $S_\infty(\tau_Q)$ for large enough $L$. In Panel B, we fit
this asymptotic value of entropy with the function
$S_{\infty}(\tau_Q)=A+B\ln\tau_Q$. The best fit has $B=0.128\pm 0.004$
and $A=1.80\pm 0.05$. This $B$ is in reasonably good agreement with the
expected value of $B=\frac{\ln 2}{12}=0.120$. In Panels C and D, we rescale
values of entropy $S(L,\tau_Q)$ by the best fit to its asymptotic value
$S_\infty(\tau_Q)=A+B\ln\tau_Q$. With this rescaling we can focus
on how the entropy depends on the block size $L$. In Panel C, we also
rescale the block size by $\hat\xi=\sqrt{\tau_Q}$ and find that the
rescaled plots do not overlap exactly. However, as shown in Panel D, rescaling 
the block size $L$ by the second scale $l=\sqrt{\tau_Q}\ln\tau_Q$ 
makes the six plots overlap quite well.
}
\label{figentropy}
\end{figure}

As observed in Ref.~\cite{17b,17c,17e,18b,JStatPhys}, the entropy can be conveniently
calculated as
\bea
S(L,\tau_Q)~=~-~{\rm Tr}~\rho~\log_2\rho~=~-~{\rm Tr}~\Pi~\log_2\Pi~.
\label{SLtauQ}
\eea
In this calculation we use Eq.~(\ref{att0}) and Eqs.~(\ref{alpha}, \ref{beta}) 
but without their large $\tau_Q$ approximations. The calculation involves a
numerical evaluation of the integrals in Eqs.~(\ref{alpha},\ref{beta}) and 
numerical diagonalization of the matrix $\Pi$. Results are shown in Panel A 
of Fig.\ref{figentropy}. The entropy grows with the block size $L$ and saturates
at a finite value $S_\infty(\tau_Q)$ for large enough $L$. In Panel B,
we fit the asymptotic entropy with the linear function
$S_{\infty}(\tau_Q)=A+B\ln\tau_Q$. The simple replacement of $\xi$ by
$\hat\xi=\sqrt{\tau_Q}$ in Eq.~(\ref{SGSoff}) suggests the asymptotic
value
\be
S_{\infty}(\tau_Q)~\simeq~
\frac16\log_2\hat\xi~\simeq~
\frac{\ln 2}{12}~\ln\tau_Q~=~
0.120~\ln\tau_Q~.
\ee
Our best fit gives $B=0.128\pm 0.004$ and $A=1.80\pm 0.05$. The best $B$ is
in reasonably good agreement with the expected value of $0.120$.

In Panels C and D of the same figure, we rescale values of entropy 
$S(L,\tau_Q)$ by its asymptotic value $S_\infty(\tau_Q)\approx A+B\ln\tau_Q$. 
After this transformation we can better focus on how the entropy depends on 
the block size $L$. A simple hypothesis would be that entropy depends on $\hat\xi$ and saturates 
when $L > \hat\xi$. To check if this is true, in Panel C we also rescale the block 
size $L$ by $\hat\xi=\sqrt{\tau_Q}$ and find that while this rescaling brings plots
close to overlap, they do not overlap as well as one might have hoped. By contrast, 
as shown in Panel D, rescaling of 
the block size $L$ by $l=\sqrt{\tau_Q}\ln\tau_Q$ makes the multiple plots overlap 
quite well indeed. In conclusion, our results support the statement that
the entropy saturates at
\bea
S_\infty(\tau_Q)~\simeq~
\frac16~\log_2\sqrt{\tau_Q}~
\label{conclS}
\eea
when 
\be
L~\gg~ \sqrt{\tau_Q}\ln\tau_Q~
\ee
i.e. the entropy of a large block of spins is determined by Kibble-Zurek
dynamical correlation length $\hat\xi=\sqrt{\tau_Q}$, but the entropy
saturates when the block size is greater than the second 
scale $l=\sqrt{\tau_Q}\ln\tau_Q$.

\begin{figure}[t]
\includegraphics[width=\columnwidth,clip=true]{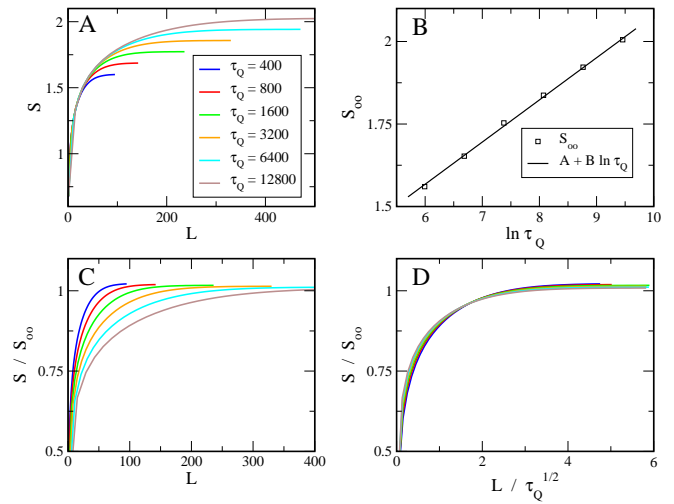}
\caption{
Panel A shows entropy of a block of $L$ spins during the dynamical phase
transition at the critical point $g=1$ as a function of the block size $L$. 
The multiple plots correspond to different values of the quench time 
$\tau_Q$. For all the quench times, the entropy grows with the block size 
$L$ and saturates at a finite value $S_\infty(\tau_Q)$ for large enough $L$. 
In Panel B we fit this asymptotic value of entropy with the function
$S_{\infty}(\tau_Q)=A+B\ln\tau_Q$. The best fit has $B=0.126\pm 0.005$
and $A=0.80\pm 0.03$. This $B$ is in reasonably good agreement with the
expected value of $B=\frac{\ln 2}{12}=0.120$. In Panels C and D, we rescale
the entropy $S(L,\tau_Q)$ by the best fit to its asymptotic value
$S_\infty(\tau_Q)\approx A+B\ln\tau_Q$. With this rescaling we can focus
on how the entropy depends on the block size $L$. In Panel D, rescaling 
the block size $L$ by $\hat\xi=\sqrt{\tau_Q}$ makes the six
plots overlap quite well.
}
\label{figentropy-g1}
\end{figure}

We believe that the KZ correlation length $\hat\xi$ is determined when 
the system is crossing the critical point while the second longer scale 
builds up after the system gets excited from its adiabatic ground state 
near the critical value of a magnetic field. At the origin of the second
scale is the non-trivial dispersion relation of excited quasiparticles. 
The $k$-dependent $\epsilon_k$ in Eq.~(\ref{epsilonk}) leads to a gradual 
evolution of matrix elements of the correlator $\Pi$ which are given
by integrals over $k$ in Eqs.~(\ref{alpha},\ref{beta}). To support this 
scenario we calculated the entropy of entanglement at the moment when 
the system is crossing the critical point at $g=1$. The results collected 
in Figure \ref{figentropy-g1} are consistent with our expectation that
near the critical point, when the scale $l$ set up by quantum phase ordering 
only begins to build up, $\hat\xi$ is still the only relevant scale of length. 

\subsection{ Impurity of the state after transition }

We were not able to do a fully analytic calculation of entropy. This
is why it may be worthwhile to calculate analytically another more 
easily tractable entanglement-related quantity. For example, the
``impurity'' of the correlator matrix $\Pi$
\be
I(\Pi)~=~ {\rm Tr}~\Pi~(1-\Pi)
\ee
is zero only when the $L$ spins are in a pure state i.e. when all 
eigenvalues of $\Pi$ are either $0$ or $1$. It is maximal when all 
the eigenvalues are $\frac12$, or when the state is most entangled. 
Thanks to its simple quadratic form, it can be calculated relatively 
easily. 

Simple calculation using the block structure of $\Pi$ in Eq.~(\ref{rhoblock}) 
and the Toeplitz property of the block matrices $\alpha$ and $\beta$ leads to
\bea
I\left(L\right) ~=~ 
2
\left( 
L\alpha_{0} - \sum_{j=1-L}^{j=L-1} (L-|j|) (\alpha_j^2 + |\beta_j|^2)      
\right),
\eea
where $\alpha_j = \alpha_{j,0}$ and $\beta_j = \beta_{j,0}$. $\alpha_j$ 
and $\beta_j$ can be expressed by the inverse Fourier transforms in Eqs.
(\ref{alpha},\ref{beta}). Using normalization $|u_k|^2+|v_k|^2=1$
and completeness of the Fourier basis we notice 
that
\bea
\alpha_0 &=& \frac{1}{2\pi}\int_{-\pi}^\pi dk~|u_k|^2 = \nonumber\\
&& \frac{1}{2\pi}\int_{-\pi}^\pi dk~(|u_k|^4+|u_kv_k^*|^2) =\\
&& \sum_{j=-\infty}^{j=\infty}(|\beta_j|^2 + |\alpha_j|^2). \nonumber
\eea
This leads to
\be
I\left(L\right) = 4L \sum_{j=L}^{\infty}(|\beta_j|^2 + |\alpha_j|^2) + 
                  4  \sum_{j=1}^{L-1}j(|\beta_j|^2 + |\alpha_j|^2).
\ee
Since we assume that $\tau_Q\gg 1$, we can approximate these sums with 
integrals. Further calculation gives:
\bea
I\left(L\right) &=& \frac{1}{2} + 
  \frac{1}{\pi}\left(1-e^{-\frac{(L-1)^2}{4 \pi \tau_Q}}\right)+ \nonumber \\
&& \frac{\ln(\tau_Q)}{2\pi^2}
   \left(1-e^{-\frac{\pi (L-1)^2}{2 \tau_Q (\ln \tau_Q)^2}}\right)^2-\frac{1}{\pi \sqrt{2 \tau_Q}}+  \nonumber \\
&& \frac{L}{2 \pi \sqrt{\tau_Q}}
   \left[\mathrm{Erfc}\left(\frac{L}{\sqrt{4\pi\tau_Q}}\right)+ \right.\\
&& \left. \sqrt{2}\mathrm{\mathrm{Erfc}}
   \left(\frac{L\sqrt{\pi}}{\sqrt{2 \tau_Q}\ln\tau_Q}\right)-
         \mathrm{Erfc}\left(\frac{L\sqrt{\pi}}{\sqrt{\tau_Q}\ln\tau_Q}\right)\right]. 
\nonumber
\eea
Here $\mathrm{Erfc}(x)$ is the complementary error function defined as:
\be
\mathrm{Erfc}(x) = \frac{2}{\sqrt{\pi}}\int_{x}^{\infty}e^{-t^2}dt.
\ee

This impurity saturates at $I_{\infty}\simeq\ln\tau_Q$ when the block size
$L\gg l$, or in short
\be
I_\infty~\approx~
\frac{\ln\tau_Q}{2\pi^2}~.
\label{Iinfty}
\ee
The impurity saturates at the second scale $l=\sqrt{\tau_Q}\ln\tau_Q$ in consistency
with our results for the entropy. 

It is interesting to compare the dynamical impurity (\ref{Iinfty}) with 
the
impurity in the ground state of the system. Simple calculation gives the asymptote 
of impurity in the ground state at the critical point
\be
I^{\rm GS}(L)~=~
\frac{\ln L}{\pi^2}~.
\ee 
when $\ln L\gg1$. Near the critical point, the asymptote is valid for the block size $L$
much less than the correlation function $L\ll\xi$ and at larger $L$ the
impurity saturates at 
\be
I^{\rm GS}_\infty~\simeq~
\frac{\ln\xi}{\pi^2}~.
\ee
Simple replacement of $\xi$ in this equation by the 
dynamical KZ correlation length $\hat\xi=\sqrt{\tau_Q}$ gives 
the asymptotic value of the dynamical impurity in Eq.~(\ref{Iinfty}).
Again, this ``replacement rule'' is the same as for the entropy.

\section{ Correlation functions }\label{correlations}

Correlation functions are of fundamental interest in phase transitions because 
they provide direct manifestation of their universal properties and are in general
easily accesible experimentally. In this Section we present our results for 
spin-spin correlation functions during a dynamical quantum phase transition.

To begin with, we observe that for symmetry reasons the magnetization 
$\langle \sigma^z \rangle=0$, but the transverse magnetization
\bea
\langle \sigma^x_n \rangle~=~ 
\langle 1-2c^\dagger_nc_n \rangle~=~
2\alpha_0-1~\approx~
\frac{1}{2\pi\sqrt{2\tau_Q}}~,
\eea
which is valid when $\tau_Q\gg 1$. This is what remains of the initial 
magnetization $\langle \sigma^x_n \rangle=1$ in the initial ground state 
at $g\to\infty$. As expected, when the linear quench is slow, then
the final magnetization decays towards $\langle \sigma^x_n \rangle=0$
characteristic of the ground state at the final $g=0$. 

Final transverse spin-spin correlation function at $g=0$ is 
\bea
C^{xx}_R &\equiv& 
\langle \sigma^x_n \sigma^x_{n+R} \rangle-  
\langle \sigma^x_n     \rangle
\langle \sigma^x_{n+R} \rangle~=~              
\label{CxxR}\\
&&
4\left(|\beta_R|^2-|\alpha_R|^2\right)~\approx \nonumber\\
&&
\frac{e^{-\frac{\pi R^2}{2~l^2}}\left(1-e^{-\frac{\pi R^2}{4~l^2}}\right)}
     {\pi~\hat\xi~l}-\frac{e^{-\frac{R^2}{\pi~\hat\xi^2}}}{2\pi^2~\hat\xi^2}~, 
\eea
when $R>1$ and $\ln\tau_Q\gg 1$. This correlation function depends on both
$\hat\xi$ and $l$. Long range correlations
\be
C^{xx}_R ~\sim~ e^{-\frac{\pi R^2}{2~l^2}} 
\ee
decay in a Gaussian way on the scale $l$.

\subsection{Ferromagnetic correlations at $g=0$}

In contrast, the ferromagnetic spin-spin correlation function 
\be
C^{zz}_R~=~
\langle \sigma^z_n \sigma^z_{n+R} \rangle~-~  
\langle \sigma^z_n \rangle
\langle \sigma^z_{n+R} \rangle~=~  
\langle \sigma^z_n \sigma^z_{n+R} \rangle            
\ee
cannot be evaluated so easily. As is well known, in the ground state, $C^{zz}_R$ 
can be written as a determinant of an $R\times R$ Toeplitz matrix whose asymptote
for large $R$ can be obtained with the Szego limit theorem \cite{Schogo}. 
Unfortunately, in time-dependent problems the correlation function is not a determinant 
in general. However, below we avoid this problem in an interesting range of parameters. 

Using the Jordan-Wigner transformation, $C^{zz}_R$ can be expressed as 
\be
C^{zz}_R ~=~ \langle b_0 a_1 b_1 a_2 \ldots b_{R-1} a_R   \rangle ~.
\label{Czzstring}
\ee
Here $a_n$ and $b_n$ are Majorana fermions defined as $a_n = (c_n^{\dagger}+c_n)$ 
and $b_n = c_n^{\dagger}-c_n$. Using (\ref{alpha}) and (\ref{beta}) we get:
\bea
\langle a_m b_n  \rangle &=& 2\alpha _ {n-m}+ 2 \Re \beta _{n-m} - \delta_{m,n}   \nonumber \\ 
\langle b_m a_n  \rangle &=& \delta_{m,n} - 2\alpha_ {n-m}+ 2 \Re \beta_{n-m}   \\ 
\langle a_m a_n  \rangle &=& \delta_{m,n} + 2 \imath \Im\beta_{m-n}  \nonumber \\
\langle b_m b_n  \rangle &=& \delta_{m,n} + 2 \imath \Im\beta_{m-n}  \nonumber 
\eea
The average in Eq.~(\ref{Czzstring}) is a determinant of a matrix when
$\langle a_m a_n  \rangle=0$ and $\langle b_m b_n \rangle=0$ for $m\neq n$, or
equivalently when $\Im\beta_{m-n}=0$ for $m\neq n$. Inspection
of the last line in Eq.~(\ref{beta}) shows that $\Im\beta_{m-n}\approx 0$ 
when $|m-n|\ll l$. Consequently, when the correlation distance $R\ll l$ 
we can neglect all $\Im\beta_{m-n}$ assuming that $ \langle a_m a_n  \rangle  = 0 $ 
and $ \langle b_m b_n  \rangle  = 0 $ for $m\neq n$. In this regime, the correlation 
function is a determinant of the Toeplitz matrix
\be
\left[ \langle b_m a_{n+1} \rangle  \right]_{m,n=1,\ldots,R} ~. 
\ee 
Asymptotic behavior of this Toeplitz determinant can be obtained using
standard methods \cite{Schogo} with the result that
\be
C^{zz}_R ~\sim~ 
\exp\left(-0.174~ \frac{R}{\hat\xi}   \right)~
\cos\left(  \sqrt{\frac{\log 2}{2 \pi}}~ \frac{R}{\hat\xi} - \varphi \right)
\label{Czz}
\ee
when $1 \ll R \ll l$. 

In this way we find that the final ferromagnetic correlation function at 
$g=0$ exhibits decaying oscillatory behavior on length scales much less 
than the phase - ordered scale $l$, but
both the wavelength of these oscillations and their exponentially decaying
envelope are determined by $\hat\xi$. As discussed in a similar situation by Cherng and Levitov
\cite{Levitov}, this oscillatory behavior means that consecutive kinks are approximately
anticorrelated -- they keep more or less the same distance $\simeq\hat\xi$ from each other 
forming something similar to a ...-kink-antikink-kink-antikink-... 
lattice with a lattice constant $\simeq\hat\xi$. However, fluctuations 
in the length of bonds in this lattice are comparable to the average 
distance itself giving the exponential decay of the correlator $C^{zz}_R$ 
on the same scale of $\simeq\hat\xi$. 

We do not know the tail of the ferromagnetic correlation function when $L\gg l$ 
because our approximations necessary to derive Eq.~(\ref{Czz}) do not work in this 
regime, but we can estimate that this tail is not negligible. Indeed, when $R=l$, 
then the envelope in Eq.~(\ref{Czz}) is
\be
\exp\left(-0.174~ \frac{l}{\hat\xi} \right)~=~
e^{-0.174\ln\tau_Q}~=~
\tau_Q^{-0.174}~.
\ee 
Due to the smallness of the exponent $0.174$ the tail is negligible only for 
extremely large $\tau_Q$.

\subsection{ Correlations at the critical point }

In order to fill in the gaps in our analytic knowledge of the correlation functions,
we attempted to make numerical simulations of the dynamical transition. As we
wanted to get information on spin-spin correlation functions it proved convenient
to work directly with spin degrees of freedom rather than with the Jordan-Wigner
fermions. We used the translationally invariant version of the real time Vidal 
algorithm \cite{Vidal}. This algorithm, which is an elegant version of the density 
matrix renormalization group \cite{DMRG}, is an efficient way to simulate time 
evolution of an {\it infinite} translationally invariant spin chain. This ambitious 
task is made efficient by a clever truncation of Schmidt decomposition between any 
two halves of the infinite spin chain. Our calculations of the entropy of entanglement 
demonstrate that for a finite transition rate, the entropy saturates at a finite value 
beyond certain block size $L$. This saturation suggests that in our case the truncation
of the Schmidt decomposition will make sense and, in principle, dynamical phase 
transitions across quantum critical points can be efficiently simulated with the 
Vidal algorithm. 

\begin{figure}[t]
\includegraphics[width=\columnwidth,clip=true]{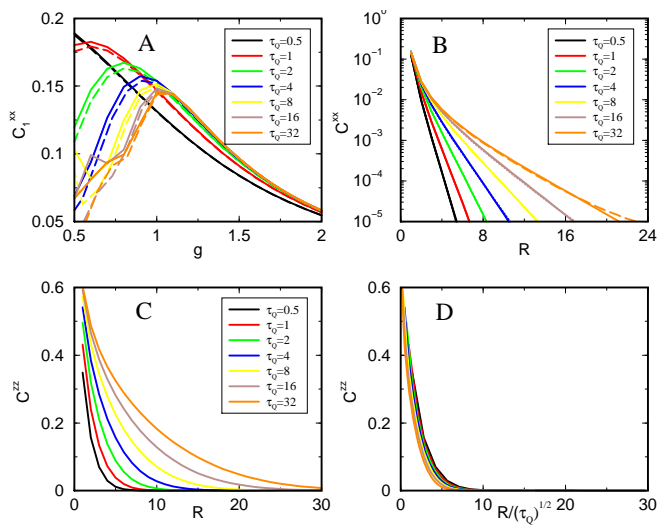}
\caption{
Panel A shows the dynamical transverse correlation $C^{xx}_1$ as a function 
of magnetic field $g$ in the linear quench. For each $\tau_Q$, we show both 
numerical (dashed) 
and analytical (solid) result. The plots overlap near the critical point at 
$g=1$ but diverge in the ferromagnetic phase when $g<1$ indicating a breakdown 
of our numerical simulations in this regime. Panel B shows analytic and numerical 
results for the dynamical transverse correlation function at the moment 
when the quench crosses the critical point at $g=1$. The transverse correlators 
overlap well confirming that our numerical simulations are still accurate 
at the critical point. Finally, in panel C, we show the dynamical ferromagnetic
correlation function $C^{zz}_R$ at $g=1$ and in panel D, we show the same
correlation function after rescaling $R/\sqrt{\tau_Q}$. The rescaled plots
overlap quite well supporting the idea that near the critical point the KZ correlation
length $\hat\xi=\sqrt{\tau_Q}$ is the only relevant scale of length.}   
\label{figVidal}
\end{figure}

In our simulations, we started the linear quench from the ground state at $g=10$
which was prepared with the imaginary time version of the algorithm. The simulations
were run for a range of $\tau_Q$ such that the initial part of the evolution
close to $g=10$ was well in the adiabatic regime. We checked our results for convergence
with respect to the truncation of the Schmidt decomposition (we used $\chi$ up to $\chi=40$)   
and time step $dt$. We used fourth order Trotter decomposition. Wherever it was possible, 
we compared our numerical results with analytical results which could be obtained
for transversal magnetization, transversal spin-spin correlations, and ferromagnetic
nearest-neighbor correlations. We also controlled if our truncation of the Schmidt
decomposition is sufficient to preserve the norm of the state evolved in real time.
As illustrated in panel A of Figure \ref{figVidal}, our simulations were stable enough 
to cross the critical point and enter the ferromagnetic phase, but once in the 
ferromagnetic phase, the algorithm was breaking down. This is why we trust our numerical 
results at $g=1$, but have no reliable results below $g=1$. We can verify KZM
at the critical point, but we cannot reliably follow the phase ordering in the 
ferromagnetic phase.

In panel B of Figure \ref{figVidal}, we plot the transverse spin-spin correlation
$C^{xx}_R$ at $g=1$ for several values of $\tau_Q$. For each $\tau_Q$, we plot 
both numerical correlator and its analytic counterpart from Eq.~(\ref{CxxR}) 
and they seem to approximately coincide. Equation (\ref{CxxR}) can be also
used to obtain analytically, but with some numerical integration, the exponential
tail of the transverse correlator when $\tau_Q\gg 1$:
\bea
C^{xx}_R 
&\approx&
\frac{0.44}{\tau_Q}~
\exp\left(-2.03 \frac{R}{\hat\xi} \right)
\label{CxxlargetauQ}
\eea 
accurate when $R\gg\hat\xi$. This tail decays on the KZ correlation length 
$\hat\xi$ which proves to be the relevant scale of length.

Encouraged by the agreement in transverse correlations in panel C we show the 
ferromagnetic spin-spin correlation functions at $g=1$ for the same values of 
$\tau_Q$. They are roughly exponential and their correlation length seems to be
set by $\hat\xi=\sqrt{\tau_Q}$. To verify this scaling hypothesis we show in 
panel D the same plots as in panel C but with $R$ rescaled as $R/\hat\xi$. We 
find the rescaled plots to overlap reasonably well confirming the expected 
$\sqrt{\tau_Q}$ scaling. The overlap is not perfect, but the scaling is expected 
when $\tau_Q\gg 1$ which is not quite satisfied by the $\tau_Q$ available from 
our numerical simulations.

\section{Conclusion}\label{conclusion}

Putting our analytical results and numerical evidence together, we are led to conclude that,
in a quantum phase transition, the system initially follows adiabatically its 
instantaneous ground state. This adiabatic behavior becomes impossible sufficiently near 
the critical point: When crossing the critical regime the system gets excited in a manner 
consistent with KZM, and imprinted with the characteristic KZ dynamical correlation length 
$\hat\xi=\sqrt{\tau_Q}$. We find evidence for this correlation length both in correlation 
functions and in the entropy of entanglement - they are all determined by the same single 
length scale $\hat\xi$. 

Once the system is excited, the non-trivial dispersion relation of its quasiparticle 
excitations leads to a gradual quantum phase ordering: Thanks to this post-critical evolution, 
the state of the system develops the second, longer, phase-ordered 
length scale which finally at $g=0$ becomes $l=\sqrt{\tau_Q}\ln\tau_Q$. This process makes short
range ferromagnetic correlation function oscillatory rather than purely exponential, 
which means that on length scales shorter than $l$ the random -kink-antikink-kink-antikink- 
train looks more like a regular crystal lattice. At the same time, thanks to 
phase ordering,
a longer block of spins is necessary to saturate the entropy of entanglement. 

It is important to note that the first process depends on the universal characteristics 
of (quantum or classical) second order phase transitions and can be modeled by KZM. 
Therefore, we expect that conclusions 
we have reached for the specific case of the quantum Ising model are generally applicable: 
Once the universality class of the transition is characterized by means of the relevant 
critical exponents, predictions of e.g. the entanglement entropy left in the wake of the 
phase transition can be made. By contrast, the dynamics of the phase ordering that follows 
can be model-specific, and is unlikely to be captured by the
scalings of relaxation time and healing length that suffice for KZM.

\section{Acknowledgements}

We thank Fernando Cucchietti and Bogdan Damski for stimulating and enjoyable discussions.
Work of L.C., J.D. and M.R. was supported in part by Polish government
scientific funds (2005-2008) as a research project and in part by 
Marie Curie ToK project COCOS (MTKD-CT-2004-517186).


\end{document}